\title{Discontinuities in the Evolution of Properties of Kerr Black Holes in the Extremal Limit}

\newcommand{\abstractText}{\noindent
Ever since its discovery by Roy Kerr in 1963, the geometry around rotating, electrostatically-neutral black holes, otherwise known as Kerr black holes, has significantly contributed to theoretical developments in the fields of general relativity, thermodynamics and beyond. Extremal Kerr black holes, in which the spin parameter of the rotating black hole is at a maximum without breaching certain postulates, has especially been of interest to the research community due to its sensitivity to higher order corrections to Einstein's theory of relativity. In this paper, we review some unresolved open questions regarding the discontinuity of certain properties of the Kerr black hole as it approaches the extreme limit. These include the form of the innermost stable circular orbit (ISCO), the nature of vanishing entropy, and the disappearance of the trapped surface as a Kerr black hole approaches extremality. We conclude that current theories are likely not suitable for studies of extremal black holes, and we posit that further development in theory may likely require a quantum approach to gravity. $G=c=1$ and Einstein summation convention are assumed throughout this paper.}

\author{Tongyu Zhang$^{(1)}$, Moe Vali$^{(2)}$\\
\textit{\textsuperscript{(1)}howardzhang0620@outlook.com; Tianjin Fayao School, China}\\
\textit{\textsuperscript{(2)}mv487@cam.ac.uk; Cavendish Laboratory, University of Cambridge, UK}}
\date{\vspace{-5ex}}

\documentclass[10pt, a4paper, tightenlines]{article}
\usepackage{xurl}
\usepackage{abstract}

\usepackage{amsmath}
\usepackage{amsfonts}
\usepackage{mathrsfs}
\usepackage{mathtools}
\usepackage{graphicx}
\usepackage{multicol}
\usepackage{caption}
\usepackage[font=small,labelfont=bf]{caption}
\usepackage{subcaption}
\usepackage{float}
\usepackage{braket}
\usepackage{cite}

\usepackage{hyperref}

\usepackage{geometry}
\begin{document}

\twocolumn[
  \begin{@twocolumnfalse}
    \maketitle
    \begin{abstract}
      \abstractText
      \newline
      \newline
      \newline
    \end{abstract}
  \end{@twocolumnfalse}]

\tableofcontents
\section{Introduction}
Black holes, formed as a consequence of supernovae which represent potential endpoints in the life cycle of the most massive of stars, are regions of spacetime separated from infinity by an event horizon \cite{carroll_spacetime_2014}. A black hole's angular momentum is largely derived from the rotation of its collapsing source star as it proceeds to a supernova. The black hole may also be electrostatically charged, but in this instance it would be swiftly neutralized by attracting matter of the opposite charge into the hole. This rotating, uncharged black hole is known as a Kerr black hole. 

The geometry around the Kerr black hole is described by the Kerr metric, which in Boyer-Lindquist coordinates is represented as the following \cite{teukolsky_kerr_2015}

\begin{align}
\begin{split}
ds^2& =-\left(1-\frac{2Mr}{\rho^2}\right)dt^2-\frac{4Marsin^2\theta}{\rho^2}d{\phi}dt +\frac{\rho^2}{\Delta}dr^2+ \\&\rho^2d\theta^2+\frac{sin^2\theta}{\rho^2}\left[\left(r^2+a^2\right)^2-a^2{\Delta}sin^2\theta\right]d\phi^2
\end{split}
\end{align}

Where $\Delta(r)=r^2-2Mr+a^2$, 
 $\rho^2(r,\theta)=r^2+a^2cos^2\theta$, and $a=J/M$ is angular momentum per unit mass. 

By inspection, one can deduce that there are two singularities in the metric, one of which occurs at 
 $\Delta=0$ and the other occurs at $\rho=0$. Through the analysis of the various scalar quantities one can make by contracting the Riemann curvature tensor for the metric, it is possible to show that the $\Delta=0$ singularity is merely a coordinate singularity and can be mitigated simply by switching to a more appropriate coordinate system. The coordinate singularities represent the event horizons of the Kerr black hole (analogous to $r=2M$ in the Schwarzschild case) \cite{teukolsky_kerr_2015}. The $\rho=0$ singularity is an intrinsic singularity (analogous to $r=0$ in Schwarzschild case). Thus, to find the event horizons, we utilise the $\Delta=0$ to find the inner horizon $r_{-}$and outer horizon $r_{+} $of the black hole to lie at 

\begin{equation}
    r_{\pm}=M\pm\sqrt{M^2-a^2}  
\end{equation}

From this expression, it is apparent that if $M>a$, then there there are two distinct event horizons. If $M<a$, then the event horizons vanish and we are left with a naked singularity, something deemed non-physical by the weak cosmic censorship conjecture \cite{zhao_testing_2024,shaymatov_higher_2019,landsman_penroses_2022}. Hence, it is impossible for a black hole to have $a>M$. In 1974, Kip Thorne proved that the spin of Kerr black holes would be limited to $a=0.998M$ due to the black hole preferentially swallowing photons of negative angular momentum emitted by the accretion flow \cite{kunde_kip_2021,benson_maximum_2009}. While recent papers have discussed the possibility of the existence of black holes beyond this limit \cite{benson_maximum_2009,aretakis_observational_2024,arbey_evolution_2020}, it remains valid that in practice $a$ can never truly reach $M$ without breaching thermodynamic laws (see \hyperlink{section.7}{Section 7}). In the theoretical case where $a=M$, the two event horizons merge into one, otherwise known as an extremal Kerr black hole. The differences in the geometry of spacetime around extremal and non-extremal Kerr black holes are contrasted in the conformal diagrams in Fig. 1 below.

\begin{figure}[h]
    \centering
    \includegraphics[width=9cm]{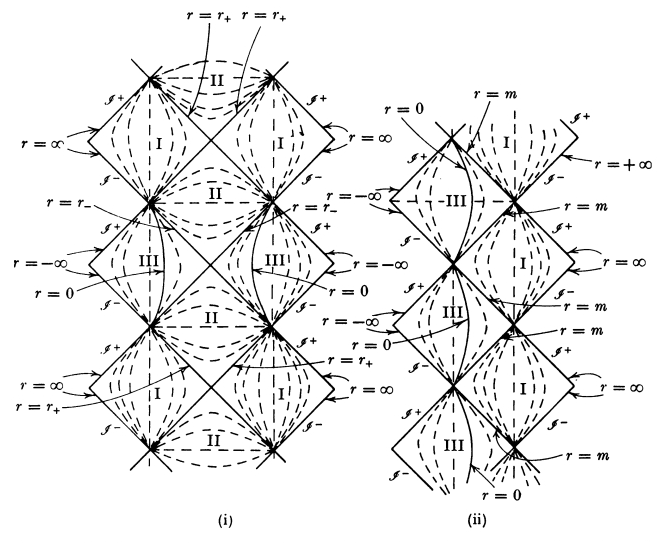}
    \caption{Conformal diagrams showing (i) the regions of spacetime in the vicinity of non-extremal Kerr black holes, and; (ii) the regions of spacetime in the vicinity of extremal Kerr black holes. Here, dotted lines represent lines of constant $r$. Region I is asymptotically flat, Region II contains trapped surfaces (see \protect\hyperlink{subsection.4.1}{Section 4.1}), and Region III contains the singularity. The two horizons $r_{+}$ and $r_{-}$ can be seen to merge into one horizon at $r=m$ as shown in (ii). Sourced from \cite{hawking_large_2023}.}
    \label{fig:enter-label}
\end{figure}

Near-extremal Kerr black holes are defined as rotating black holes which have the property $(r_{+}-r_{-})/r_{1}<<1$, where $r_{1}$ is the $r$ coordinate of the event horizon in the extremal case of a black hole with identical mass \cite{pradhan_extremal_2013}.

Extremal black holes serve as a very useful theoretical tool for testing higher-order corrections to Einstein's theory of relativity, a theory considered to be a low energy approximation to a more complete theory \cite{horowitz_extremal_2023,silveravalle_gravitational_2022,gabadadze_new_2023}. This is because it can be shown that generic stationary, axisymmetric
perturbations generate singularities on the horizon, and when one extends the near-horizon geometry to a full asymptotically flat solution one necessarily introduces these singular perturbations \cite{horowitz_extremal_2023}. Hence, extremal black holes are sensitive to higher-order corrections to Einstein's theory, and in this paper a discussion of potential failures of Einstein's theory of general relativity are highlighted in the form of discontinuities of certain Kerr black hole properties in the near-extremal limit. These concerns may be effectively addressed by theories of quantum gravity.

\addtocontents{toc}{\protect\setcounter{tocdepth}{2}}
\section{ISCO of extremal vs. near-extremal Kerr black holes}

\subsection{Extremal case}
In this subsection, we analyze the ISCO of the extremal Kerr black hole and establish its correspondence to null geodesics on the event horizon with $d\Tilde{\phi}/d\Tilde{t}=1/2M$, following the work presented in \cite{pradhan_extremal_2013}. However, the validity of this correspondence remains unclear in the case of the near-extremal Kerr black hole (see \hyperlink{subsection.2.2}{Section 2.2}). Because the Boyer-Lindquist coordinate system is divergent as $\Delta$ approaches 0, we make use of the Doran coordinate system yielding the Kerr metric \cite{jacobson_where_2011} 
\begin{align}
\begin{split}
    ds^2& =-d\Tilde{t}^2+\left[ \sqrt{\frac{2Mr}{\rho^2}}(d\Tilde{t}-asin^2{\theta}d\Tilde{\phi})+\sqrt{\frac{\rho^2}
    {r^2+a^2}}dr\right]^2\\
   & +\rho^2d\theta^2+\left(r^2+a^2\right)sin^2{\theta}d\Tilde{\phi^2}
\end{split}
\end{align}
which are related with the Boyer-Lindquist coordinates by
\begin{align}
    d\Tilde{t}=dt+\frac{\beta}{1-\beta^2}dr
\end{align}
\begin{align}
    d\Tilde{\phi}=d\phi+\frac{a\beta}{\left(1-\beta^2\right)\left(r^2+a^2\right)}dr
\end{align}
 Here $\rho$ follows the same definition as in the Boyer-Lindquist form and $\beta$ is defined by
 \begin{align}
    \beta=\sqrt{\frac{2Mr}{r^2+a^2}}
\end{align}
 Restricting ourselves to orbits in the equatorial plane where $\theta=\pi/2$, this results in the metric
\begin{align}
\begin{split}
    ds^2& =-\left(1-\frac{2M}{r}\right)d\Tilde{t^2}-\frac{4M^2}{r}d\Tilde{t}d\Tilde{\phi}+2\sqrt{\frac{2Mr}{r^2+M^2}}\Tilde{dt}dr-\\& 2M\sqrt{\frac{2Mr}{r^2+M^2}}drd\Tilde{\phi}+\frac{r^2}{r^2+M^2}dr^2+\\&
     \left(r^2+M^2+\frac{2M^3}{r}\right)d\Tilde{\phi}^2
\end{split}
\end{align}
The metric is independent of the variables $\Tilde{t}$ and $\Tilde{\phi}$, and hence $p_{\Tilde{\phi}}$ and $p_{\Tilde{t}}$ are conserved along geodesics where $\vec{p}$ is the four-momentum of the test particle \cite{schutz_first_2018}. Defining the conserved quantities $\mathcal{E}=-p_{\Tilde{t}}$ and $L=p_{\Tilde{\phi}}$ yields
\begin{align}
    \mathcal{E}=\left(1-\frac{2M}{r}\right)p^{\Tilde{t}}-\sqrt{\frac{2Mr}{r^2+M^2}}p^r+\frac{2M^2}{r}p^{\Tilde{\phi}}
\end{align}
\begin{align}
L=-\frac{2M^2}{r}p^{\Tilde{t}}-M\sqrt{\frac{2Mr}{r^2+M^2}}p^r+\left(r^2+M^2+\frac{2M^3}{r}\right)p^{\Tilde{\phi}}
\end{align}
Analyzing circular orbits on the horizon where $r=M$ and $p^r=0$, it is possible to show that $L=2M\mathcal{E}$ \cite{pradhan_extremal_2013}. The scalar created when the four-momentum is contracted with itself at $r=M$ yields

\begin{align}
\vec{p}\cdot\vec{p}=\mathcal{E}^2
\end{align}
If $\mathcal{E}=0$, then it follows that $L=0$ and this represents null geodesics on the horizon of the extremal black hole. Using $\mathcal{E}=0$, it follows that the angular velocity  $\Omega$ of the particle is given by
\begin{align}
    \Omega=\frac{d\Tilde{\phi}}{d\Tilde{t}}=\frac{p^{\Tilde{\phi}}}{p^{\Tilde{t}}}=\frac{1}{2M}
\end{align}
\subsection{Near-extremal case}
Summarising the analysis for non-extremal Kerr geometry in the same coordinate system and using the results presented in \cite{pradhan_extremal_2013}, one obtains
\begin{align}
\mathcal{E}=\left(1-\frac{2M}{r}\right)p^{\Tilde{t}}-\sqrt{\frac{2Mr}{r^2+a^2}}p^r+\frac{2aM}{r}p^{\Tilde{\phi}}
\end{align}
\begin{align}
L=-\frac{2aM}{r}p^{\Tilde{t}}-a\sqrt{\frac{2Mr}{r^2+a^2}}p^r+\left(r^2+a^2+\frac{2Ma^2}{r}\right)p^{\Tilde{\phi}}
\end{align}
For circular orbits $p^r=0$. An expression for $p^{\Tilde{t}}$ and $p^{\Tilde{\phi}}$ can be obtained by application of Cramer's rule. One can then prove that
\begin{align}\label{14}
\vec{p}\cdot\vec{p}=-\mathcal{E}p^{\Tilde{t}}+Lp^{\Tilde{\phi}}
\end{align}

For circular orbits at $r=r_0$, it is possible to recast Eq. 14 in the following form
\begin{align}
\vec{p}\cdot\vec{p}=\frac{1}{\Delta}\left[(L^2-a^2\mathcal{E}^2)-r_0^2\mathcal{E}^2-\frac{2M}{r_0}(L-a\mathcal{E})^2\right]
\end{align}
For the extremal limit, we apply $a{\rightarrow}M$. Given we are in the vicinity of the event horizon, we apply $\Delta=(r_0-M)^2\rightarrow0$. Taking these limits, we obtain \cite{pradhan_extremal_2013}
\begin{align}
\vec{p}\cdot\vec{p}=-\frac{(L-2M\mathcal{E})^2}{(r_0-M)^2}
\end{align}
A difficulty arises when we insert the exact extremal case i.e. $L=2M\mathcal{E}$ and $r_0=M$, into Eq. 16, leading to ambiguous interpretations for whether we obtain the true null geodesics from the exactly extremal case given both numerator and denominator are concurrently zero. 
\subsection{Reconciling the discontinuity}
As presented by \cite{pradhan_extremal_2013}, there is no agreement of the ISCO in the limiting case of the near-extremal Kerr black hole with the exactly extremal case. The limit may however be considered to be continuous, as suggested by \cite{gralla_particle_2015} where it is stated that orbits at $r=r_0$ possess the property (to leading order)
\begin{align}
    \frac{\Omega-\Omega_H}{\Omega_H}=-\frac{3(r_0-r_{+})}{4r_{+}}
\end{align}
Where $\Omega_H$ is the horizon angular velocity, defined by 
\begin{align}
\Omega_H=\frac{a}{a^2+r_{+}^2}
\end{align}
as shown in \cite{carroll_spacetime_2014}. It can be seen that $\Omega_H$ is a continuous function. For orbits at the outer horizon $r_0=r_{+}$ Eq. 17 vanishes so $\Omega=\Omega_H$ is also continuous.

One possible approach in reconciling this discontinuity involves considering the event horizon as a Killing horizon. In this approach, the event horizon is defined as a hypersurface where the Killing vector is null.

In a stationary and non-static spacetime such as the region outside the outer horizon of a Kerr black hole, the event horizon is also the Killing horizon for the Killing vector $\chi^{\mu}$ defined in the Boyer-Lindquist coordinate system by 
\begin{align}
\chi^{\mu}=K^{\mu}+{\Omega}_HR^{\mu}
\end{align}
as discussed in \cite{carroll_spacetime_2014}. Here $K^{\mu}=(\partial_t)^{\mu}$, $R^{\mu}=(\partial_{\phi})^{\mu}$, with $\Omega_H$ a constant. It has been shown that for the Kerr metric, $\Omega_H$ can be identified with the horizon angular velocity.
Transforming this expression into Doran coordinates, we obtain
\begin{align}
    \chi^{\mu'}=\frac{{\partial}x^{\mu'}}{{\partial}t}+{\Omega_H}\frac{{\partial}x^{\mu'}}{{\partial}\phi}
\end{align}
where primed and unprimed coordinates represent Doran coordinates and Boyer-Lindquist coordinates, respectively. Using Eq. 4 and 5, the only modifications required to make Eq. 19 valid in Doran coordinates is a change of $R^{\mu}$ from $({\partial_{\phi}})^{\mu}$ to $({\partial}_{\Tilde{\phi}})^{\mu}$ and a change of $K^{\mu}$ from $({\partial_{t}})^{\mu}$ to $({\partial_{\Tilde{t}}})^{\mu}$.

Null surfaces such as Killing horizons possess the property of being unable to have two linearly independent null tangent vectors \cite{carroll_spacetime_2014}. This means the four-momentum of the particle, a null tangent vector to one of the geodesics that make up the generator for the null hypersurface, must be a multiple of the Killing vector $\vec{\chi}$. From this, we can conclude that
\begin{align}
    \Omega=\frac{d\Tilde{\phi}}{d{\Tilde{t}}}=\frac{d\Tilde{\phi}/d\lambda}{d\Tilde{t}/d\lambda}=\frac{p^{\Tilde{{\phi}}}}{p^{\Tilde{t}}}=\frac{{\chi}^{\Tilde{\phi}}}{{\chi}^{\Tilde{t}}}=\Omega_H
\end{align}
Where $\lambda$ is some affine parameter. For extremal Kerr black holes the circular null geodesic on the horizon has $\Omega=1/2M$ (\hyperlink{subsection.2.1}{Section 2.1}), which is exactly $\Omega_H$ in the extremal case, making this equation valid for the exactly extremal case. Eq. 21 is continuous, yielding the same result in the limiting case of near-extremal black holes, thereby potentially mitigating the issue of discontinuity.

\section{Entropy of black holes}
\subsection{Origin of entropy in black holes}
At first review, it may be considered counter-intuitive for black holes to have non-zero entropy. According to statistical mechanics $S=kln\Omega$ where $S$ is the entropy, $k$ is Boltzmann's constant, and $\Omega$ is the number of microstates associated with a particular macrostate \cite{blundell_concepts_2010}. According to the no-hair theorem of black holes \cite{ruffini_introducing_1971} recently proven to be true for isolated, asymptotically flat, stationary, axisymmetric black holes (such as Kerr black holes in $f(R)$ theories of gravity) \cite{sultana_no-hair_2018}, the characteristics of a black hole are fully determined by its mass, charge, and angular momentum. Hence, one may naively conclude that black holes have no entropy as there is only one configuration of charge, mass, and angular momentum that could create a particular black hole. It should however be noted that the no-hair theorem may not be valid in alternate theories of gravity, as discussed in \cite{brihaye_skyrmions_2017,kanti_dilatonic_1998,bekenstein_black_1996}.  

If there is a single time-independent field without event horizons, such as the case of a hypothetical naked singularity, then there is only one configuration of stress energy that could generate the field, and this would correspond to zero entropy \cite{edery_extremal_2011}. However, it is important to consider that the black hole must have been formed by the collapse of matter, and the contents of that matter are shielded from the outside observers by the non-stationary region (non-stationary meaning all Killing vectors are spacelike) between the inner and outer horizons. This region is present for all non-extremal black holes, including Reissner-Nordstrom black holes and Kerr-Newman black holes \cite{carroll_spacetime_2014}. The contents of the material inside this region cannot be determined outside the hole, and this generates the microstates that give the black hole its entropy. 

According to Bekenstein, the entropy of black holes originates from the ignorance of the internal configurations of the black hole to outside observers \cite{edery_extremal_2011,bekenstein_black_1973}. We may therefore expect that the entropy of a black hole may in fact be quite large; the black hole could be made from anything and arranged in any shape, pattern, or size, as  long as its net charge, net angular momentum, and total mass parameters are valid. In the Bekenstein approach, a black hole of one solar mass would have an entropy of approximately 18 orders of magnitude larger than the thermodynamic entropy of the Sun in our solar system \cite{bekenstein_black_1973}.

\subsection{Absence of entropy in extremal case}
It has been proven that the entropy of a black hole is proportional to the area of its event horizon, with the proportionality constant being $1/4\hbar$ in units where $G=c=1$ \cite{lousto_fourth_1993,strominger_microscopic_1996,carlip_black_2014,howard_geometric_2013}. This is true for black holes close to the extremal limit, but many research papers have pointed out that this equation breaks down for the exactly extremal black hole \cite{edery_extremal_2011,carroll_extremal_2009,belgiorno_black_2004}. This is to be expected, as in the extremal limit the two event horizons merge into one, and the non-stationary region, which is what gives the black hole its entropy in the first place, vanishes. Here, we outline a review of how classical results break down, reviewing the work by Howard \cite{howard_geometric_2013}. Suppose the entropy of the black hole is given by some arbitrary function of the area of an event horizon
\begin{align}
S_{BH}=f(A)
\end{align}
If the area of the event horizon is perturbed by some small amount ${\delta}A$, then f(A) is changed by some amount ${\delta}f(A)$, given by
\begin{align}
{\delta}f(A)=\frac{df(A)}{dA}{\delta}A
\end{align}
This change in event horizon area must have been caused by an alteration in the internal configurations of the black hole, and this could occur if matter falls into the black hole. Suppose a particle were to fall into the black hole. After it crosses the outer horizon, there is no information about the particle that can be obtained for outside observers. Hence, the particle could either still exist or not exist with equal probability, and this gives two possible microstates to the macrostate. Thus, the minimum entropy change is \cite{howard_geometric_2013}
\begin{align}
({\delta}f(A))_{min}=kln(\Omega)=kln2
\end{align}
For Kerr black holes, the first law of thermodynamics reads \cite{johnstone_extremal_2013}
\begin{align}
    dM=TdS+\Omega_HdJ
\end{align}
Which for small perturbations in black hole mass, entropy and angular momentum can be recast in the following form
\begin{align}
    {\delta}M-\Omega_H{\delta}J=T{\delta}S
\end{align}
The black hole's event horizon area will change accordingly by \cite{carroll_spacetime_2014}
\begin{align}
    {\delta}A=\frac{8{\pi}a({\delta}M-\Omega_H{\delta}J)}{{\Omega_H}\sqrt{M^2-a^2}}
\end{align}
As extremal black holes have zero temperature \cite{carroll_spacetime_2014}, the expression in Eq. 26 is equal to zero, and hence the numerator in Eq. 27 also equals zero. However, the denominator in Eq. 27 vanishes too, and so the expression for ${\delta}A$ is undefined. Hence, no conclusions can be drawn for the value of $df(A)/dA$ indicating a failure of the classical approach.

\subsection{Beyond semi-classical theories}
It is pertinent to note that while the Bekenstein entropy equation can be proven under certain assumptions \cite{carroll_spacetime_2014}, there are many alternate modern theories regarding the entropy of black holes that extend beyond semi-classical methods. Some theories of quantum gravity explain the entropy by attributing it to the quantum field correlations between the exterior and interior of the black hole \cite{carroll_extremal_2009}. Others relate the black hole entropy to the ordinary entropy of its thermal atmosphere \cite{wald_thermodynamics_2001}. Attempts have also been made to explain the entropy of black holes in the context of Sakharov’s theory of induced gravity, in which the dynamical aspects of gravity arise from the collective excitations of massive fields \cite{wald_thermodynamics_2001}. More recent approaches involve the framework of quantum geometry \cite{wald_thermodynamics_2001}. The holographic principle attributes the entropy to the encoding of the degrees of freedom in a gravitating system to its spatial boundaries \cite{wall_survey_2018}. String theorists have also made attempts to explain the entropy by using methods of microstate counting. It has been found that a string theory approach gives a non-vanishing entropy for exactly extremal black holes, and gives a value which agrees with the classical prediction of the value of entropy calculated when taking the near-extremal limit of non-extremal black holes \cite{carroll_extremal_2009,wald_thermodynamics_2001}. For the We consider extremal black holes to have no entropy, as there are far more research papers which hold this viewpoint \cite{howard_geometric_2013,belgiorno_black_2004,edery_extremal_2011,lousto_fourth_1993}. 

\section{The disappearance of the trapped surface in the extremal limit
}
\subsection{Trapped surfaces}

A trapped surface is defined by a compact two-dimensional submanifold embedded in original manifold such that the expansion for all future directed light rays is negative everywhere on the submanifold \cite{carroll_spacetime_2014}. 

\begin{figure}[h]
    \centering
    \includegraphics[width=6cm]{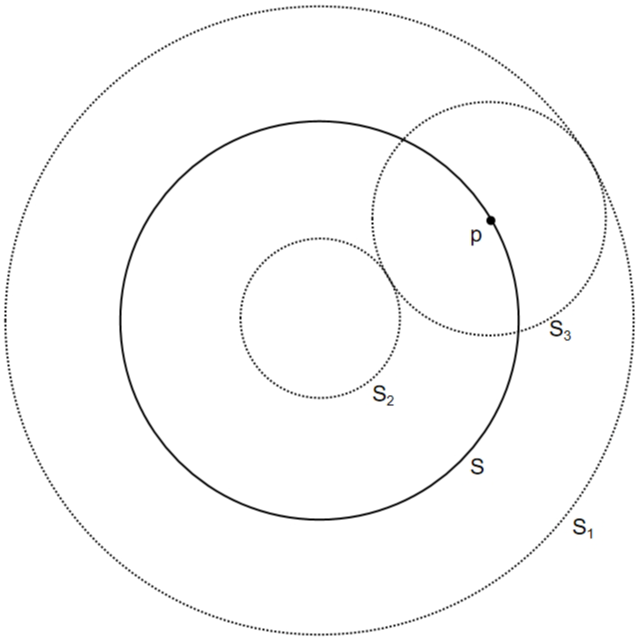}
    \caption{Overview of the definition of trapped surfaces, adapted from \cite{hawking_large_2023}. Here a trapped surface is defined as a compact two-dimensional submanifold embedded in original manifold. See text for a further description.}
    \label{fig:enter-label}
\end{figure}

The expansion $\theta$ is defined by
\begin{align}
\theta=\nabla_{\mu}{q}^{\mu}
\end{align}
where ${q}^{\mu}=dx^{\mu}/d\lambda$ is the null vector representing the light ray. Light rays with this property are said to converge \cite{carroll_spacetime_2014}. From Fig. 2, consider a sphere $S$ which emits a flash of light. At some later time, the light emitted from a point $p$ on the sphere $S$ would form a sphere $S_3$ which touches the wavefront of the ingoing light $S_2$ and outgoing light $S_1$. If the expansion is less than zero, then the surface area of the wavefront will be smaller than S, and if both the ingoing and outgoing wavefronts satisfy this condition then matter inside, which cannot travel faster than light, are trapped inside the surface whose area falls to zero within a finite time \cite{hawking_large_2023}. 

\subsection{Trapped surfaces in Kerr spacetime}
For these calculations, we work with the metric induced by the original metric on the spatial slices generated by the null geodesic as time progresses, defined by \cite{pradhan_extremal_2013}
\begin{align}
\gamma_{{\mu}{\nu}}=g_{{\mu}{\nu}}+l_{\mu}n_{\nu}+n_{\mu}l_{\nu}
\end{align}
Where $l^{\mu}$ represent an ingoing future directed null vector and $n^{\mu}$ represents an outgoing one. Here $g_{{\mu}{\nu}}$ is the metric in Boyer-Lindquist coordinates. Of course, we could scale the vectors by arbitrary constant scalars and the vectors would still represent future directed null vectors, and hence for the sake of simplicity we choose the two vectors such that 
\begin{align}
l^{\mu}n_{\mu}=-1
\end{align}
Then the two null rays can be represented as follows \cite{pradhan_extremal_2013}
\begin{align}
l^{\mu}=\frac{1}{\Delta}(r^2+a^2,-\Delta,0,a)
\end{align}
\begin{align}
n^{\mu}=\frac{1}{2r^2}(r^2+a^2,\Delta,0,a)
\end{align}
Using this, one can calculate the expansion for both these null vectors. They are \cite{pradhan_extremal_2013}
\begin{align}
\theta_{(l)}=-\frac{2}{r}
\end{align}
\begin{align}
\theta_{(n)}=\frac{\Delta}{r^3}
\end{align}
$\theta_{(l)}$ always converges, but this is to be expected as it is ingoing. The ingoing null ray emanating into spacetime from the surface of a two-sphere embedded in Minkowski spacetime converges too. Both of these are negative for $r_{-}<r<r_{+}$, which indicates the existence of trapped surfaces. However, if the black hole is extremal then $r_{-}=r_{+}$ and this region would vanish. This means $\Delta$ is never negative and so there is no trapped surface in the extremal case. This can also be seen from Fig. 1 as Region II in diagram (ii) vanishes.

\section{The problem}

Most concepts in physics involve the concept of differentiability, something lost if discontinuities are present. Consequently, having discontinuities in certain physical properties such as entropy or the form of the ISCO could potentially violate known physics. In classical statistical mechanics, entropy is a function of state which can be derived from the partition function, a continuous function \cite{blundell_concepts_2010}. In the classical theory of general relativity, Einstein's field equation link the energy-momentum of a system to it's spacetime curvature in a continuous manner, which would not permit a sudden change in the form of the ISCO as $a{\rightarrow}M$. The disappearance of the trapped surface further shows the continuity issues with viewing extremal Kerr black holes as the limiting case of near-extremal Kerr black holes. These discontinuities in properties spark consequences which violate known physics, and hence present an open problem in the realm of theoretical physics. 

\section{An optimistic answer}
\subsection{New theories?}
General relativity is one of the simplest possible metric theories of gravity, and it has successfully withstood all experimental tests conducted so far. The equation that link the metric tensor with the stress energy tensor is known as Einstein's field equation, and one way to derive it is through the principle of least action. In n dimensions, the equation for the action S is given by \cite{carroll_spacetime_2014}
\begin{align}
S=\int{\mathcal{L}({\Phi^i,{\nabla_{\mu}{\Phi}}^i}})d^nx
\end{align}
where $\mathcal{L}$ is the Lagrange density, and ${\Phi}^i(x)$ a set of fields which act as the dynamical variables. In general relativity, the dynamical variable is the metric $g_{{\mu}{\nu}}$, and it is necessary to construct a scalar from the metric that could serve as the Lagrange density \cite{carroll_spacetime_2014}. Since the laws of general relativity permits us to analyze local problems in local Lorentz frames where the equations of special relativity can be used \cite{schutz_first_2018}, it follows that metric can be set to it's canonical form and first derivatives all set to zero at any point in the manifold. Hence, any nontrivial scalar must have at minimum second derivatives of $g_{{\mu}{\nu}}$. Naturally, the Riemann curvature tensor comes to mind. It can be shown that any nontrivial tensor made from products of the metric, it's derivatives, and it's second derivatives can be expressed in terms of the metric and the Riemann curvature tensor \cite{carroll_spacetime_2014}. The Ricci scalar R is a rank 0 tensor made from products of the metric, it's derivatives, and it's second derivatives. It is also the only independent scalar one could construct from the Riemann curvature tensor \cite{carroll_spacetime_2014}. Hence, the only scalar choice for the Lagrange density which is no higher than second order in the derivatives of the metric tensor is the Ricci scalar. Technically, multiples would also suffice, but they don't affect how we minimize the action. The Hilbert action $S_H$ is given by
\begin{align}
S_H=\int{\sqrt{-g}Rd^nx}
\end{align}
Where $g$ represents the determinant of the metric tensor. It can be shown that minimizing this function yields Einstein's equation in vacuum \cite{carroll_spacetime_2014}.

But why is it acceptable to only consider terms no larger than second order? What about terms like $R^2$, $R_{{\mu}{\nu}}R^{{\mu}{\nu}}$, ${\nabla}^{\mu}R{\nabla}_{\mu}R$? Why rule them out? While no experiments have revealed these terms, theories regarding quantum gravity have suggested the existence of these terms \cite{silveravalle_gravitational_2022,capozziello_higher-order_2000,velasco-aja_quantum_2022}. For example, one problem that has plagued Einstein's theory is that when one attempts to consistently quantize general relativity, one discovers that the theory is not renormalizable. As it turns out, this obstacle can be removed if one adds the right combination of higher-order Lagrange densities to the original Lagrange density \cite{carroll_spacetime_2014,bezerra-sobrinho_modified_2023}.

The author holds the view that perhaps the non-physical discontinuities one encounters when taking the extremal limit of a black hole could be due to our current theory of relativity being incomplete, as it is just a classical theory and not a quantum one. After all, the discontinuity problem of entropy vanishes if one analyzes the problem from the point of view of microstates using string theory, and as mentioned earlier in the paper these black holes are incredibly sensitive to higher-order corrections to Einstein's theory. So it isn't too far fetched to believe that corrections from string theory or quantum physics can be the key to solving the other problems too. Also, while discontinuities are indeed non-physical in classical theories, they occur all the time in quantum physics (for example the possible eigenvalues of the Hamiltonian operator in bound states). It is entirely possible that when we look back at this problem in the future equipped with a more complete theory of quantum gravity these problems will fade away. 

\subsection{Where to look}
In particular, I believe that string theory is most likely to resolve the discrepancies of the discontinuites that occur in the extremal limit. Many theorems in classical physics, for example the equipartition theorem, require that the thermal energy is much larger than the energy gap between quantized energy levels so that quantum effects can be washed out \cite{blundell_concepts_2010}. This is clearly not the case for extremal black holes as they have vanishing temperature, so it follows that to model extremal black holes quantum theories must be used. Hence, it comes as no surprise that Einstein’s classical theory of relativity breaks down. What we are searching for then is a theory that can unify quantum physics and gravitation, and this is precisely what string theory is. 

String theory can fix many problems in general relativity, like for example the discontinuity in black hole entropy in the extremal limit and the problems that occur when one attempts to quantize Einstein's theory \cite{zwiebach_first_2009}. There are other reasons to believe that string theory is a good unified theory of quantum physics and gravity. For example, there are no adjustable parameters in string theory, as opposed to in the Standard Model of particle physics where there are about twenty \cite{zwiebach_first_2009}. The absence of adjustable parameters mean we know for a fact that researchers aren’t just adjusting the parameters to fit other known theories, giving any confirmations to the theory more value. There is an abundance of research papers on the topic of black holes that make use of string theory with many different approaches to explain the physics behind them (see \cite{myers_black_2001} for a list of approaches), and the author believes that there is a reason to this. The author would suggest any future researchers who want to delve into the topic of the discontinuity of properties of Kerr black holes in the extremal limit to look into string theory, as the author believes that this theory is most likely to bear fruit. 

\section{A pessimistic answer}

Another possibility is that it is not Einstein's theory that is incomplete, but merely just that an unrealistic system is analyzed. It has been proven that black holes cannot reach $a=M$ as mentioned previously in the paper. To further prove that extremal black holes cannot form, the author devised a thought experiment of his own.

We propose a hypothetical situation where a  black hole has just formed. Initially it would not be extremal as extremal black holes have zero temperature, which according to the third law of black hole thermodynamics (The weaker version of the third law of thermodynamics, to which black holes obey) cannot be obtained within a finite amount of steps \cite{socolovsky_black_2023,carlip_black_2014} and hence cannot form as a result of matter collapse. To increase it's angular momentum per unit mass it would have to change it's internal configurations. One way it could do this is by absorbing matter with positive angular momentum to increase the value of angular momentum per unit mass. Suppose at some instance it is very very close to extremality. At this instant a tennis ball is thrown at it which has just the right mass and angular momentum to turn it into an extremal Kerr black hole. Initially, the black hole and tennis ball would both have entropy, but after the tennis ball falls in the black hole becomes extremal and the entropy vanishes. This is a clear violation of the generalized second law, which states that the net entropy of matter and black holes can never decrease in an isolated system \cite{carlip_black_2014,lousto_fourth_1993}.

Another way it could change it's internal configurations is by emitting Hawking radiation. Suppose again we have a black hole that is very very close to extremality, and by emitting Hawking radiation it reduces it's mass just enough such that $a/M=1$ and it becomes extremal. Before the radiation is emitted, the black hole has entropy, but after the radiation is emitted the entropy vanishes, again violating the generalized second law of thermodynamics. Hence, we conclude that for black holes  close to extremality Hawking radiation must affect it's physical parameters in a way such that it evolves away from extremality, as also suggested by other studies \cite{di_gennaro_how_2022}. As neither the emission nor the absorption of particles can generate an extremal black hole without breaching the generalized second law of thermodynamics, it is clear that extremal Kerr black holes are non-physical.

Utilizing general relativity to study extremal Kerr black holes can be likened to employing special relativity to examine scenarios where the object's velocity exceeds the speed of light. In such cases, peculiar conclusions may arise, such as the Lorentz factor becoming imaginary, but that's not because the theory is flawed, it's because an impossible system is being analyzed. So it is also possible that the non-physical discontinuities coming up in the theory of extremal Kerr black holes is merely because we are dealing with an unrealistic system, and that further analysis of it may not be of much value. 

\section{Conclusion}

This paper has shown that evidently the details of the theory of extremal black holes is still lacking, as there are many non-physical discontinuities in the evolution of the physical properties of the black hole as it approaches extremality. First, It is revealed that while the ISCO of the extremal black hole can be shown to be represented by null geodesics on the horizon of the black hole, it remains unclear whether this holds true in the extremal case. We then went on to show that while non-extremal black holes have entropy that obey the Bekenstein-Hawking formula, the entropy for extremal black holes vanish although it's event horizon area does not, leading to a disagreement with the formula and another discontinuity. Finally, we showed that the trapped surface disappears when a black hole reaches extremality.

As of March 2024, there are no concrete explanations for these discontinuities. Of course, it could just be because a non-physical system is analyzed, but the author is more tempted to believe that extremal black holes could hold the key to new, potentially revolutionizing physics. Gravity is the weakest of the four fundamental forces, and hence little is known about it's effects on small scales where usually other forces such as electromagnetic forces dominate. In the extreme conditions of black holes, gravity is the force that dominates over the other forces, and this could lead to failure of some of our classic models and would require a quantum theory to explain. Progress is currently being made by these non-classical theories to explain the black hole's entropy \cite{carroll_extremal_2009,de_haro_conceptual_2020,das_black_1996,dorey_black_2023,nomura_black_2014}, and it may just be a matter of time before experts in the field develop a more complete theory where all the discontinuity problems vanish, and we will then look back at these problems in the same way we view the ultraviolet catastrophe today. Not only will this help us better understand black holes, but will also be helpful in analysis of other areas where gravity is strong such as the interior of neutron stars and the properties of the universe during the very first moments of the big bang. In that sense, extremal black holes could act as a testing site for these new theories, as they are very sensitive to higher order corrections of spacetime curvature. Indeed, the possibilities are boundless, and the further research can be pursued by future researchers.

\section{Data availability}
No new data was used in this paper.

\section{Conflicts of interest}
The authors declare no conflicts of interest.

\bibliography{bibtex}

\begin{thebibliography}{10}

\bibitem{carroll_spacetime_2014}
S.~M. Carroll, {\em Spacetime and geometry: an introduction to general relativity}.
\newblock Harlow: Pearson Education, pearson new international edition~ed., 2014.

\bibitem{teukolsky_kerr_2015}
S.~A. Teukolsky, ``The {Kerr} {Metric},'' {\em Classical and Quantum Gravity}, vol.~32, p.~124006, June 2015.
\newblock arXiv:1410.2130 [astro-ph, physics:gr-qc, physics:hep-th].

\bibitem{zhao_testing_2024}
M.~Zhao, M.~Tang, and Z.~Xu, ``Testing {The} {Weak} {Cosmic} {Censorship} {Conjecture} in {Short} {Haired} {Black} {Holes},'' Feb. 2024.
\newblock arXiv:2402.16373 [gr-qc].

\bibitem{shaymatov_higher_2019}
S.~Shaymatov, N.~Dadhich, and B.~Ahmedov, ``The higher dimensional {Myers}-{Perry} black hole with single rotation always obeys the {Cosmic} {Censorship} {Conjecture},'' {\em The European Physical Journal C}, vol.~79, p.~585, July 2019.
\newblock arXiv:1809.10457 [astro-ph, physics:gr-qc, physics:hep-th].

\bibitem{landsman_penroses_2022}
K.~Landsman, ``Penrose's 1965 singularity theorem: {From} geodesic incompleteness to cosmic censorship,'' {\em General Relativity and Gravitation}, vol.~54, p.~115, Oct. 2022.
\newblock arXiv:2205.01680 [gr-qc, physics:math-ph, physics:physics].

\bibitem{kunde_kip_2021}
H.~Kunde, ``On {Kip} {Thorne}{\textquoteright}s limit for rotation of black holes,'' 2021.

\bibitem{benson_maximum_2009}
A.~J. Benson and A.~Babul, ``Maximum {Spin} of {Black} {Holes} {Driving} {Jets},'' {\em Monthly Notices of the Royal Astronomical Society}, vol.~397, pp.~1302--1313, Aug. 2009.
\newblock arXiv:0905.2378 [astro-ph].

\bibitem{aretakis_observational_2024}
S.~Aretakis, G.~Khanna, and S.~Sabharwal, ``An observational signature for extremal black holes,'' Jan. 2024.
\newblock arXiv:2307.03963 [gr-qc, physics:hep-th, physics:math-ph].

\bibitem{arbey_evolution_2020}
A.~Arbey, J.~Auffinger, and J.~Silk, ``Evolution of primordial black hole spin due to {Hawking} radiation,'' {\em Monthly Notices of the Royal Astronomical Society}, vol.~494, pp.~1257--1262, May 2020.
\newblock arXiv:1906.04196 [astro-ph, physics:gr-qc].

\bibitem{hawking_large_2023}
S.~Hawking, G.~F.~R. Ellis, and A.~Ashtekar, {\em The large scale structure of space-time}.
\newblock Cambridge monographs on mathematical physics, Cambridge, United Kingdom: Cambridge University Press, 50th anniversary edition~ed., 2023.

\bibitem{pradhan_extremal_2013}
P.~Pradhan and P.~Majumdar, ``Extremal {Limits} and {Kerr} {Spacetime},'' {\em The European Physical Journal C}, vol.~73, p.~2470, June 2013.
\newblock arXiv:1108.2333 [gr-qc].

\bibitem{horowitz_extremal_2023}
G.~T. Horowitz, M.~Kolanowski, G.~N. Remmen, and J.~E. Santos, ``Extremal {Kerr} black holes as amplifiers of new physics,'' {\em Physical Review Letters}, vol.~131, p.~091402, Aug. 2023.
\newblock arXiv:2303.07358 [astro-ph, physics:gr-qc, physics:hep-ph, physics:hep-th].

\bibitem{silveravalle_gravitational_2022}
S.~Silveravalle, ``The gravitational field of isolated objects in quadratic gravity,'' {\em Il Nuovo Cimento C}, vol.~45, pp.~1--4, June 2022.
\newblock arXiv:2202.00999 [gr-qc, physics:hep-th].

\bibitem{gabadadze_new_2023}
G.~Gabadadze, ``A {New} {Gravitational} {Action} {For} {The} {Trace} {Anomaly},'' {\em Physics Letters B}, vol.~843, p.~138031, Aug. 2023.
\newblock arXiv:2301.13265 [gr-qc, physics:hep-ph, physics:hep-th].

\bibitem{jacobson_where_2011}
T.~Jacobson, ``Where is the extremal {Kerr} {ISCO}?,'' {\em Classical and Quantum Gravity}, vol.~28, p.~187001, Sept. 2011.
\newblock arXiv:1107.5081 [gr-qc, physics:hep-th].

\bibitem{schutz_first_2018}
B.~F. Schutz, {\em A {First} {Course} in {General} {Relativity}}.
\newblock Cambridge: University Press, second edition, 13th printing~ed., 2018.

\bibitem{gralla_particle_2015}
S.~E. Gralla, A.~P. Porfyriadis, and N.~Warburton, ``Particle on the innermost stable circular orbit of a rapidly spinning black hole,'' {\em Physical Review D}, vol.~92, p.~064029, Sept. 2015.

\bibitem{blundell_concepts_2010}
S.~Blundell and K.~M. Blundell, {\em Concepts in thermal physics}.
\newblock Oxford: Oxford university press, 2nd ed~ed., 2010.

\bibitem{ruffini_introducing_1971}
R.~Ruffini and J.~A. Wheeler, ``Introducing the black hole,'' {\em Physics Today}, vol.~24, pp.~30--41, Jan. 1971.

\bibitem{sultana_no-hair_2018}
J.~Sultana and D.~Kazanas, ``A no-hair theorem for spherically symmetric black holes in \${R}{\textasciicircum}2\$ gravity,'' {\em General Relativity and Gravitation}, vol.~50, p.~137, Nov. 2018.
\newblock arXiv:1810.02915 [gr-qc].

\bibitem{brihaye_skyrmions_2017}
Y.~Brihaye, C.~Herdeiro, E.~Radu, and D.~H. Tchrakian, ``Skyrmions, {Skyrme} stars and black holes with {Skyrme} hair in five spacetime dimension,'' {\em Journal of High Energy Physics}, vol.~2017, p.~37, Nov. 2017.
\newblock arXiv:1710.03833 [gr-qc, physics:hep-th].

\bibitem{kanti_dilatonic_1998}
P.~Kanti, N.~E. Mavromatos, J.~Rizos, K.~Tamvakis, and E.~Winstanley, ``Dilatonic {Black} {Holes} in {Higher}-{Curvature} {String} {Gravity} {II}: {Linear} {Stability},'' {\em Physical Review D}, vol.~57, pp.~6255--6264, May 1998.
\newblock arXiv:hep-th/9703192.

\bibitem{bekenstein_black_1996}
J.~D. Bekenstein, ``Black hole hair: twenty--five years after,'' May 1996.
\newblock arXiv:gr-qc/9605059.

\bibitem{edery_extremal_2011}
A.~Edery and B.~Constantineau, ``Extremal black holes, gravitational entropy and nonstationary metric fields,'' {\em Classical and Quantum Gravity}, vol.~28, p.~045003, Feb. 2011.
\newblock arXiv:1010.5844 [gr-qc, physics:hep-th].

\bibitem{bekenstein_black_1973}
J.~D. Bekenstein, ``Black {Holes} and {Entropy},'' {\em Physical Review D}, vol.~7, pp.~2333--2346, Apr. 1973.

\bibitem{lousto_fourth_1993}
C.~O. Lousto, ``The {Fourth} {Law} of {Black} {Hole} {Thermodynamics},'' {\em Nuclear Physics B}, vol.~410, pp.~155--172, Dec. 1993.
\newblock arXiv:gr-qc/9306014.

\bibitem{strominger_microscopic_1996}
A.~Strominger and C.~Vafa, ``Microscopic {Origin} of the {Bekenstein}-{Hawking} {Entropy},'' {\em Physics Letters B}, vol.~379, pp.~99--104, June 1996.
\newblock arXiv:hep-th/9601029.

\bibitem{carlip_black_2014}
S.~Carlip, ``Black {Hole} {Thermodynamics},'' {\em International Journal of Modern Physics D}, vol.~23, p.~1430023, Oct. 2014.
\newblock arXiv:1410.1486 [gr-qc].

\bibitem{howard_geometric_2013}
E.~M. Howard, ``Geometric aspects of {Extremal} {Kerr} black hole entropy,'' {\em Journal of Modern Physics}, vol.~04, no.~03, pp.~357--363, 2013.
\newblock arXiv:1511.00594 [gr-qc].

\bibitem{carroll_extremal_2009}
S.~M. Carroll, M.~C. Johnson, and L.~Randall, ``Extremal limits and black hole entropy,'' {\em Journal of High Energy Physics}, vol.~2009, pp.~109--109, Nov. 2009.
\newblock arXiv:0901.0931 [hep-th].

\bibitem{belgiorno_black_2004}
F.~Belgiorno and M.~Martellini, ``Black {Holes} and the {Third} {Law} of {Thermodynamics},'' {\em International Journal of Modern Physics D}, vol.~13, pp.~739--770, Apr. 2004.
\newblock arXiv:gr-qc/0210026.

\bibitem{johnstone_extremal_2013}
M.~Johnstone, M.~M. Sheikh-Jabbari, J.~Simon, and H.~Yavartanoo, ``Extremal {Black} {Holes} and {First} {Law} of {Thermodynamics},'' {\em Physical Review D}, vol.~88, p.~101503, Nov. 2013.
\newblock arXiv:1305.3157 [hep-th].

\bibitem{wald_thermodynamics_2001}
R.~M. Wald, ``The {Thermodynamics} of {Black} {Holes},'' {\em Living Reviews in Relativity}, vol.~4, p.~6, Dec. 2001.
\newblock arXiv:gr-qc/9912119.

\bibitem{wall_survey_2018}
A.~C. Wall, ``A {Survey} of {Black} {Hole} {Thermodynamics},'' July 2018.
\newblock arXiv:1804.10610 [gr-qc, physics:hep-th].

\bibitem{capozziello_higher-order_2000}
S.~Capozziello and G.~Lambiase, ``Higher-{Order} {Corrections} to the {Effective} {Gravitational} {Action} from {Noether} {Symmetry} {Approach},'' {\em General Relativity and Gravitation}, vol.~32, pp.~295--311, Feb. 2000.
\newblock arXiv:gr-qc/9912084.

\bibitem{velasco-aja_quantum_2022}
E.~Velasco-Aja and J.~Anero, ``Quantum corrections to {Einstein}'s equations,'' Sept. 2022.
\newblock arXiv:2209.06938 [gr-qc].

\bibitem{bezerra-sobrinho_modified_2023}
J.~Bezerra-Sobrinho and L.~G. Medeiros, ``Modified {Starobinsky} inflation by the \${R}{\textbackslash}ln{\textbackslash}left( {\textbackslash}square{\textbackslash}right) {R}\$ term,'' {\em Journal of Cosmology and Astroparticle Physics}, vol.~2023, p.~039, Jan. 2023.
\newblock arXiv:2202.13308 [astro-ph, physics:gr-qc].

\bibitem{zwiebach_first_2009}
B.~Zwiebach, {\em A first course in string theory}.
\newblock Cambridge ; New York: Cambridge University Press, 2nd ed~ed., 2009.

\bibitem{myers_black_2001}
R.~C. Myers, ``Black {Holes} and {String} {Theory},'' July 2001.
\newblock arXiv:gr-qc/0107034.

\bibitem{socolovsky_black_2023}
M.~Socolovsky, ``Black {Holes} and the {Third} {Law} of {Thermodynamics} {Revisited},'' {\em Journal of High Energy Physics, Gravitation and Cosmology}, vol.~09, no.~02, pp.~499--505, 2023.
\newblock arXiv:2304.07890 [gr-qc].

\bibitem{di_gennaro_how_2022}
S.~Di~Gennaro and Y.~C. Ong, ``How {Not} to {Extract} {Information} {From} {Black} {Holes}: {Cosmic} {Censorship} as a {Guiding} {Principle},'' {\em Physics Letters B}, vol.~829, p.~137112, June 2022.
\newblock arXiv:2103.05516 [gr-qc, physics:hep-th].

\bibitem{de_haro_conceptual_2020}
S.~De~Haro, J.~van Dongen, M.~Visser, and J.~Butterfield, ``Conceptual {Analysis} of {Black} {Hole} {Entropy} in {String} {Theory},'' {\em Studies in History and Philosophy of Science Part B: Studies in History and Philosophy of Modern Physics}, vol.~69, pp.~82--111, Feb. 2020.
\newblock arXiv:1904.03232 [hep-th, physics:physics].

\bibitem{das_black_1996}
S.~R. Das, ``Black {Hole} {Entropy} and {String} {Theory},'' Feb. 1996.
\newblock arXiv:hep-th/9602172 version: 1.

\bibitem{dorey_black_2023}
N.~Dorey, R.~Mouland, and B.~Zhao, ``Black {Hole} {Entropy} from {Quantum} {Mechanics},'' {\em Journal of High Energy Physics}, vol.~2023, p.~166, June 2023.
\newblock arXiv:2207.12477 [hep-th].

\bibitem{nomura_black_2014}
Y.~Nomura and S.~J. Weinberg, ``Black {Holes}, {Entropies}, and {Semiclassical} {Spacetime} in {Quantum} {Gravity},'' {\em Journal of High Energy Physics}, vol.~2014, p.~185, Oct. 2014.
\newblock arXiv:1406.1505 [gr-qc, physics:hep-th].

\end{thebibliography}

\bibliographystyle{ieeetr}

\end{document}